\documentclass[letterpaper,twocolumn,prl]{revtex4}
\usepackage[latin9]{inputenc}
\usepackage{verbatim}
\usepackage{mathtools}
\usepackage{amsmath}
\usepackage{amssymb}
\usepackage{graphicx}
\usepackage{wasysym}
\usepackage{ulem}

\makeatletter

\pdfpageheight\paperheight
\pdfpagewidth\paperwidth

\usepackage{lipsum}
\usepackage{dcolumn}
\usepackage{bm}
\usepackage{bm}
\usepackage{braket}
\usepackage{color}
\usepackage{float}
\usepackage{subfigure}
\usepackage{color}
\usepackage{soul}
\usepackage{sidecap}
\usepackage{float}

\makeatother

\begin{document}
\title{
Weak measurement based pseudo-spin pointer: a cost-effective scheme for precision measurement
}
\author{Ling Ye$^1$, Lan Luo$^{2,3}$, An Wang$^1$, Rongchun Ge$^{1*}$, Zhiyou Zhang$^{1*}$}
\affiliation{$^1$College of Physics, Sichuan University, Chengdu 610064, China}
\affiliation{$^2$Key Laboratory of Science and Technology on Space Optoelectronic Precision Measurement, CAS, Chengdu 610209, China}
\affiliation{$^3$Institute of Optics and Electronics, Chinese Academy of Sciences, Chengdu 610209, China}
\affiliation{Corresponding author: rcge@scu.edu.cn; zhangzhiyou@scu.edu.cn}
%

\begin{abstract}
As an essential component of state-of-the-art quantum technologies, fast and efficient quantum measurements are in persistent demand over time. Here, we present a precision measurement scheme for weak signals. We propose a dimensionless pseudo-spin pointer based on weak measurement which effectively converts a continuous measurement into a discrete one. In the context of optical parameter estimation, we demonstrate experimentally that a weak perturbation of the parametric distribution's moment can be retrieved efficiently by employing the dimensionless pointer without measuring the distribution literally. In addition to the sheer liberation of experimental expense, the photon-countering-based pointer is well-calibrated for the detection of weak signals. We show that for signals $3$-$4$ orders of weaker in strength than the area-array camera method, an order of improvement in precision is achieved experimentally.
\end{abstract}
\maketitle

{\it Introduction.} Ever-expanding quantum technologies have been a focus of public interests, and have routinely cultivated new research opportunities as well as challenges~\cite{q1,q2,q3,q5,q6}. It is generally believed that one of the main elements that renders the quantum advantages over its classical counterpart is the non-local coherence of a quantum system; consequently, a quantum system can be steered strategically to achieve the desired evolution in parallel. As all of the designed procedures are done at the end of operations, a typical quantum measurement needs to be carried out to retrieve the relevant information encoded in the state of the system.

Quantum measurement as an unsettled domain of quantum theory has sponsored numerous new ideas which have not only deepened our understanding of fundamental properties of the quantum world, like coherence, but also provided new measurement schemes with great precision and efficiency. Weak measurement, proposed in the late 1980s by Aharonov, Albert, and Vaidmain~\cite{q7}, has provided a generic framework to extract the information of a quantum system and amplify its numerical value not striking the system brutally. Astonishing results pop out of the scrutiny of weak measurement from time to time, which have gradually transformed our perception of the nature of quantum theory. The concept of weak value application (WVA) has been explored intensively since then~\cite{the1,the2,the3,the4}. In the following decades, the framework of WVA was implemented in a wide spectrum of experimental research~\cite{exp1,exp2,exp3,exp4}, among which was the pioneering work of Hosten {\it et al}. on spin Hall effect of light (SHEL)~\cite{q14}. The generic scheme of the pre- and post-selections of the system of interest which couples to an ancilla has been found versatile even in the regime beyond weak coupling~\cite{nonpertwm,q15}.

Precision measurements, which are essential for high technology and also of vital importance to fundamental science, have gained renewed momentum with the exploration of quantum regime. To beat the standard quantum limit implicated by the well-known central limit theorem, it seems quantum resources~\cite{q18} are needed. The quantum metrology protocol presented by Giovannetti {\it et al}.~\cite{q19,q20,q21} showed that initial entanglement among input states for otherwise independent measurement setups can be used to achieve $1/\sqrt{N}$ improvement over the standard quantum limit, namely the Heisenberg limit. Recently, interesting observations show that with additional input of correlation, super-Heisenberg is also possible~\cite{q22,q23}. In previous experiments employing non-classical photon sources such as single or entangled photons to break the standard quantum limit, an effective coupling between the relevant photons and the target system is designed proportional to the parameter(s) to be measured~\cite{q24,q25,q26,q27,q28,q29}. Mathematically, the physics of the measurement is converted to a problem of parameter estimation. As usual, in the above protocols, the statistics of the relevant (small) parameter are measured directly, and the moments of the parameter are derived from the distribution function. 

To put it more specifically, we take the measurement of position (momentum) in an optical experiment as an example. In order to obtain the mean value of position (momentum), the previous wisdom is to obtain the distribution of the position (momentum) directly from the statistics of measurement~\cite{exp1,exp2,q14,q30}. Technically, it is done with multi-pixel detectors such as charge-coupled devices (CCD) and scientific complementary metal-oxide-semiconductor (sCMOS), which have a high spatial resolution. However, the charge has to be shifted pixel by pixel and processed by amplifiers, or a large number of charge signals need to be processed simultaneously. So the reading efficiency of the detector is limited. Moreover, interference between the circuit components is easy to occur, which impacts the measuring quality even further. These issues make the manufacturing process of the register circuit extremely complicated with a low yield. As a consequence, it can be time-consuming, hardware-expensive, and high power consumption. 

In this letter, we present a dimensionless pseudo-spin pointer to achieve a cost-effective (in terms of both time and hardware) measurement of the moments of position with two single-pixel detectors such as single-photon avalanche diode (SPAD). The two SPADs consist of the pseudo-spin here, and we will take one as spin-up and the other as spin-down for convenience. It is well known that SPAD has a really bad spatial resolution, but has a low threshold and short response time which render it ideal to record the clinching of incoming photons. The key point of our measurement is to obtain the relative photon counting numbers of the two SPADs ($N_+$ for spin-up, and $N_-$ for spin-down). This working mechanism enables us to derive the information of the moment of the position through the contrast ratio of the counting numbers: $|\frac{N_+-N_-}{N_++N_-}|$. With this dimensionless pointer, we achieve the measurement of the moment of position without measuring the position! This enables us to obviate the drawback of SPAD while making full use of its merits of high sensitivity and accuracy. Consequently, our pseudo-spin pointer is naturally adapted for the measurement of weak signals.
We demonstrate that the Cram\'er-Rao bound can be achieved in practical measurement\cite{q31,q32,q33} employing the strategy here. With the experiment setup of SHEL, we show proof-of-principle verification by controlling the small variation of the post-selection angle. More sensitive and accurate experimental results are obtained by diminishing the number of pixels. The experimental results show that the precision achieved is comparable to that employing CCD with much higher spatial resolution. Although the intensity of input light is 3$\sim$4 orders of magnitude weaker than the working condition of CCD, an improvement of one order for the accuracy of the output signal is obtained. Our current measurement scheme offers a serious choice when dealing with extremely weak signals or when imminent feedback is in need. 


\begin{figure}[htbp]
\centering
\includegraphics[width=1\linewidth]{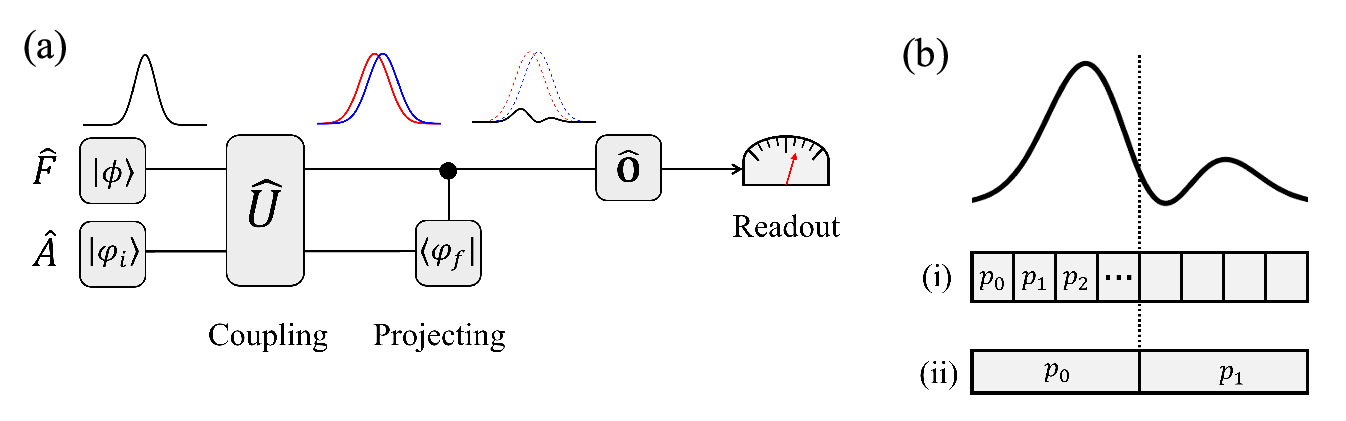}
\caption{(a) Schematic diagram of the method based on the WVA experiment.  (b) Comparison between the conventional processing method and the dimensionless pseudo-spin pointer method. By acquiring the probabilities over the two regions, the pointer can decode the information about the weak coupling and post-selection.}
\label{fig:01}
\end{figure}
{\it Theory and experimental proposal.} 
To obviate any plausible ambiguity from using the notation, we will give a brief summary of the WVA experiment. As illustrated in Fig. \ref{fig:01}(a), the meter is prepared in a pure state $|\phi\rangle$ and the system is initialized with a pure pre-selection state as $|\varphi_{i}\rangle$. The evolution of the composite system of the meter and the system is described by $\hat{\text U}=\exp(-{\rm ig}\hat{\rm A}\otimes\hat{\rm F})$, where $\hat{\rm A}$ is an observable of the system, $\hat{\rm F}$ is an observable for the meter and ${\rm g}$ is the effective weak-coupling strength. Final projection on the system is carried out by selecting a post-selection state $|\varphi_{f}\rangle$. Effectively, the evolution of the meter state can be described by $|\Phi\rangle=\hat{\rm M}|\phi\rangle$, where the Kraus operator $\hat{\rm M}=\langle\varphi_{f}|\hat{\rm U}|\varphi_{i}\rangle$ describes the effect of the parametric interaction of the system on the initial state $|\phi\rangle$. For any observable $\hat{\rm O}$ of the meter, its averaged numerical value is given by $\langle\hat{\rm O}\rangle= \langle\phi|\hat{\rm M}^{\dagger}\hat{\rm O}\hat{\rm M}|\phi\rangle/ \langle\phi|\hat{\rm M}^{\dagger}\hat{\rm M}|\phi\rangle$.

For concreteness, we assume the system is pre-selected as a two-level system in the eigenbasis of $\hat{\rm A}$: $|\varphi_{i}\rangle=(|a_{1}\rangle+|a_{2}\rangle)/
{\sqrt2}$, where $\hat{\rm A}=|a_{1}\rangle\langle a_{1}|-|a_{2}\rangle\langle a_{2}|$. The post-selection state is intentionally chosen as $|\varphi_{f}\rangle={\rm i} (e^{-{\rm i}\theta}|a_{2}\rangle-e^{{\rm i}\theta}|a_{1}\rangle)/{\sqrt2}$, which is approximately orthogonal to the initial state $|\varphi_i\rangle$.

The main input of the new measurement scheme comes from dividing the Hilbert space into two disjoint subspaces, which defines the pseudo-spin degree of freedom
\begin{equation}\label{eq2}
\begin{aligned}
|\phi\rangle&=\frac{1}{\sqrt2}(|0\rangle+|1\rangle)\\
&\equiv\int_{0}^{\infty} d q f(q)|q\rangle+\int_{-\infty}^{0} d q f(q)|q\rangle.
\end{aligned}
\end{equation}
Here $f(q)$ is a normalized wave function that encodes the probability amplitude and is assumed as a Gaussian distribution $f(q)=({2}/{\pi\sigma^2})^{1/4} \exp(-{q^2}/{\sigma^2 })$ as in our experiment following. $|0\rangle$ and $|1\rangle$ are the spin-up and spin-down of the pseudo-spin, which physically implies the record of clinching of the upper and bottom detectors at the arrival of photons. The relevant observables of the meter are $\hat{\rm F}=\hat{\rm p}$, and $\hat{\rm O}=|0\rangle\langle0|-|1\rangle\langle1|$, respectively. 

As a result, the measurement of $\hat{\rm O}$ lead to $\langle\hat{\rm O}\rangle\approx \big(2{\rm g}{\rm Im}(\alpha {\rm A}_{w})+{\rm g}^{2}\beta|{\rm A}_{w}|^{2}\big)/\big(1+\dfrac{{\rm g}^{2}}{\sigma^{2}}|{\rm A}_{w}|^{2}\big)$; 
with $\alpha=\sqrt{2/{\pi}}\sigma$, $\beta=0$, and ${\rm A}_{w}=\langle\varphi_{f}|\hat{\rm A}|\varphi_{i}\rangle/\langle\varphi_{f}|\varphi_{i}\rangle=-{\rm i}\cot\theta$ is the weak value. Then the observable turns into:

\begin{equation}\label{eq4}
\langle\hat{\rm O}\rangle=2\sqrt{\frac{2}{\pi}}\frac{\rm g}{\sigma}\frac{\cot\theta}{1+(\frac{\rm g}{\sigma}\cot{\theta})^{2}}.
\end{equation}

As explained above, $\hat{\rm O}$ is the third component of the Pauli matrix in the space of pseudo-spin, which physically renders the particle number contrast by the two detectors and represents the distribution of the meter in the coordinate space. 

Physically, this result demands only the probabilities over the two regions rather than the detailed distribution over the whole space as is shown in Fig. \ref{fig:01}(b). Compared with the previous method of detecting the light field's center of mass, this scheme can obtain the position information of the photon with the minimum number of pixels which makes it respond much faster while demanding much less. This is the key observation that underlines the merits of the new measurement scheme as explained below. In the present work, the influence of uniformly distributed technical noise can be minimized by subtracting each part of electron fluctuations in the two-state system. Moreover, by constructing a purely imaginary weak value ${\rm A}_w$, the technical noises can also be effectively suppressed~\cite{q34,q35}. In the meantime, the response of the detectors is maintained in the dynamic range because of fewer photons applied. In general, the measurement based on the dimensionless output signal (as shown below) significantly relieves the burden of deliberate photon counting and calculation. It is expected that the scheme will outperform the traditional ones and the experimental results following will show firm evidence for the claimed advantages above.

\begin{figure}[htbp]
\centering
\includegraphics[width=1\linewidth]{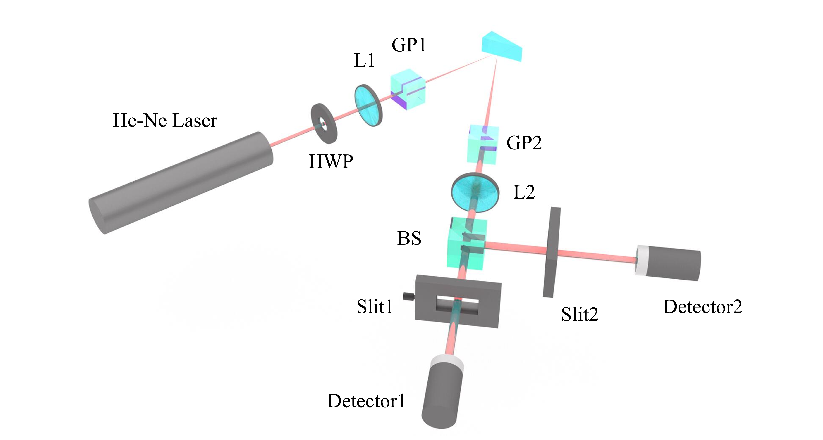}
\caption{The Experimental Setup. Light source: He-Ne laser; HWP: binary compound zero-order half-wave plate; L1 and L2: lenses form the 4f system, which can focus and collimate; GP1 and GP2: Glan Polarizers; BS: Beam Splitter; Slit1 and Slit2 separate the upper and the bottom parts of the photons. Single-photon detectors are used to collect the output photons.}
\label{fig:02}
\end{figure}

\begin{figure*}[htbp]
\centering
\includegraphics[width=1\linewidth]{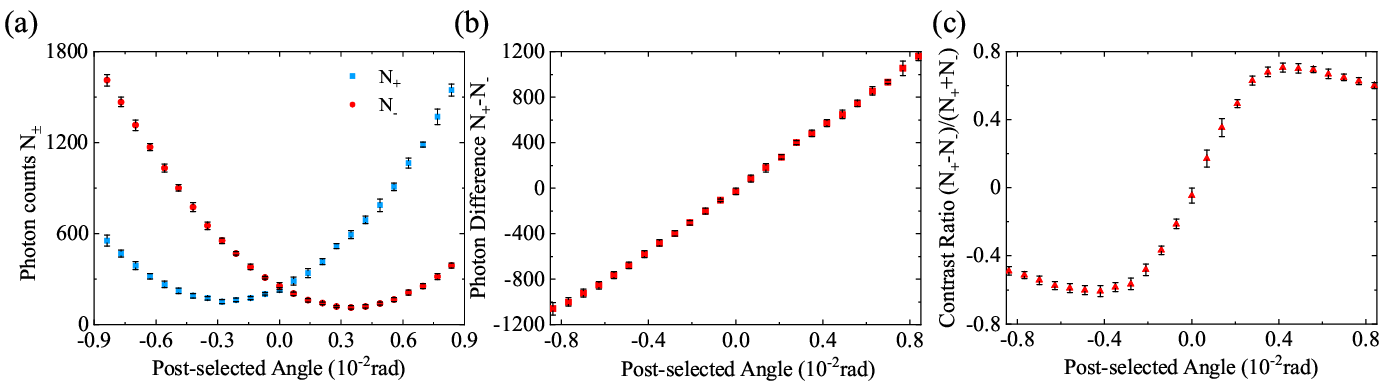}
\caption{(a), (b) show the collected photons ($N_+, N_-, N_+-N_-$) from two single-photon detectors changing with the post-selected angle. (c) shows the signal $\frac{N_{+}-N_{-}}{N_{+}+N_{-}}$ changing with the post-selected angle $\theta$.}
\label{fig:03}
\end{figure*}

As a specific case of parameter estimation, we show that the Cram\'er-Rao bound of this measurement can be saturated. In the subsequent process, $\langle\hat{\rm O}\rangle$ is detected with respect to the changing range of post-selected angle $\hat{\theta}$. The meter's associated probabilities after the post-selection are defined as $p(+|\theta) = {1}/{2}+\sqrt{{2}/{\pi}}{\rm{g} \sigma\cot\theta}/({\sigma^{2}+\rm{g}^{2}\cot^{2}\theta})$, and $p(-|\theta)  = 1-p(+|\theta)$.
For the numbers of photons ($N_+$ and $N_-$) that arrive at either side of the detector, we can get the distribution of these is $p\left( {N_{-}, N_{+}~} \right) = \left\lbrack \frac{N_{+}! + N_{-}!}{N_{-}!N_{+}!} \right\rbrack{p(+ |\theta)}^{N_{+}}{p(- |\theta)}^{N_{-}}$.
$N=N_+ + N_-$ represents the total number of photons after the post-selection. The final signal $\frac{N_+-N_-}{N_++N_-}$ is the physical meaning of $\langle \hat{\rm O} \rangle$. So we can obtain the variance of $\hat{\rm O}$ is ${\rm Var}[\hat{\rm O}]={4}p(+|\theta)p(-|\theta)/{N}$. Furthermore, the slope of the signal is obtained by $\partial \langle\hat{O}\rangle/{\partial \theta}$, which satisfies ${\partial p(+ |\theta)}/{\partial \theta} = ({\partial \langle\hat{\rm O}\rangle}/{\partial \theta})/2 = - {\partial p(- |\theta)}/{\partial \theta}$. Therefore, the error-propagation formula gives the estimator's sensitivity:

\begin{equation}\label{eq8}
{\rm Var}[\hat{\theta}]=\frac{{\rm Var}[\hat{\rm O}]}{|\partial \langle\hat{\rm O}\rangle/\partial \theta|^2}=\frac{1}{N F_\theta}, 
\end{equation}
where the Fisher information is $F_{\theta} = {\sum_{\pm}\frac{\left\lbrack {\partial_{\theta}p(\pm |\theta)} \right\rbrack^{2}}{p(\pm |\theta)}}$. Obviously, the inverse of ${\rm Var}[\hat{\theta}]$ is exactly the Fisher information, which verifies that the sensitivity obtained from the error-propagation formula can saturate the Cram\'er-Rao bound over the entire estimator interval.


{\it Experimental results and analysis.}  
We begin by presenting a brief description of the experimental procedure followed to estimate the variations of the unknown parameter. The parameter to be estimated is introduced by adjusting the post-selection angle $\theta$. The weak coupling between the system and the meter is accomplished by the SHEL~\cite{q14,SHEL1,SHEL2,SHEL3}.

The experimental setup is shown in Fig. \ref{fig:02}. First, the Gaussian beam is produced by a He-Ne laser with a wavelength of 632.8nm and a beam width of $\sigma =27$~$\mu$m~\cite{width}. By rotating the half-wave plate (HWP), the photons passing through the Glan polarizer 1 (GP1) are extremely attenuated to fit the detectors. The Glan polarizers (GP1 and GP2) perform the pre- and post-selections of the polarization of photons and the lenses play the roles of focus and collimation. After the pre-selection, incident photons with linear polarization can be written initialized as $\left|\left.\varphi_{i}\right\rangle\ \right.=\left(\left|\left.+\right\rangle\right.+\left|\left.-\right\rangle\right.\right)/{\sqrt2}$, where $|+\rangle$ and $|-\rangle$ represent the left- and right-handed circular polarization states, respectively. The Glan polarizer 2 is used to produce linearly polarized light with a small angle $\theta$ to the vertical polarized light. So, the post-selected state can be regarded as $|\varphi_{f}\rangle={\rm i}( e^{-{\rm i}\theta}|-\rangle-e^{{\rm i}\theta}|+\rangle)/{\sqrt2}$.

The normal SHEL takes place when the photons are reflected on the air-glass surface. A BK7 prism with the refractive index $n=1.515$ is used as the reflective plane in the actual experiment. Coupled with the momentum $k_y$, the photons undergo a small evolution, which could be described as an operator $\hat{\rm U}=exp(-{\rm i}k_y{\hat{\sigma}}_3\delta_H)$. Here, ${\hat{\sigma}}_3=|+\rangle\langle+|-|-\rangle\langle-|$ is the Pauli operator. The photons with opposite spin have different spatial shifts $\delta_H =\cot\theta_i(1+r_s/r_p)/k_0$ denotes the initial splitting due to the spin-orbit interaction, where $r_p$ and $r_s$ are the Fresnel coefficients, $k_0$ represents the vacuum wave vector and the incident angle $\theta_i$ is $30^\circ$. By rotating GP2, different contrast ratios $(N_+-N_-)/(N_++N_-)$ are acquired that correspond to the range of post-selected angles $\theta$ as

\begin{equation}\label{eq9}
\frac{N_+-N_-}{N_++N_-}=2\sqrt{\frac{2}{\pi}}\frac{{k_0} \sigma  r_p (r_p+r_s)\cot \theta_i\cot\theta}{{k_0}^2{\sigma^2{r_p}^2+({r_p+r_s})^2}{\cot\theta_i^2{\cot\theta}^2}}.
\end{equation}

\begin{figure}[htbp]
\includegraphics[width=1\linewidth]{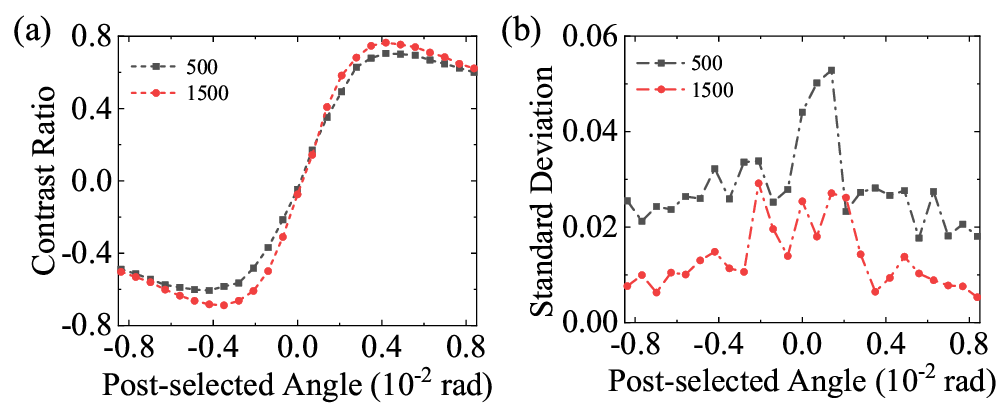}
\caption{The comparison of different output photon counts ($N_{min}$=500 and 1,500). Integration time is 10ms. (a) is the contrast ratio. To facilitate comparison, the corresponding standard deviations are shown in (b). }
\label{fig:04}
\end{figure}

In the case of extremely weak signals, CCD suffers from problems such as weak detection ability and slow response time. In sharp contrast with the ordinary measurement, the strategy of measurement based on the dimensionless pointer obviates the need for the explicit measurement of position information. As a result, at the detection port, a high sensitivity and high precision of SPAD can be employed. This bestows a great experimental advantage of our strategy over the previous method on the measurement of weak signals. The two SPADs will just measure the number of photons in the upper and lower halves respectively. The corresponding power is converted into a power unit around $10^{-5}\sim10^{-4}$nW at the receiving photon frequency of $50$~kHz (which means only 50,000 photons are counted per second). This power intensity is far below the working range of commonly used CCDs available in the experiment.
We measured multiple sets of data at different signal intensities and integration times. The number of photons received by the single-photon detectors is recorded respectively as $N_+$ and $N_-$ and the contrast ratio is obtained by subtraction and normalization ($({N_+-N_-})/({N_++N_-})$). In order to obtain a decent estimate of the standard derivation, ten sets of data at each point were measured. Detailed data processing is shown in Fig. \ref{fig:03}. It can be seen that the variation trends of the measured value $N_+$ and $N_-$ are approximately symmetric with respect to the post-selection angle $\theta=0^\circ$ (shown in Fig. \ref{fig:03}(a)). The contrast ratio is shown in Fig. \ref{fig:03}(c). It's worth noting that the trend of the contrast ratio is like previous measurements\cite{q36,q37,q38}. The signal appears as a linear response to the post-selected angle when it is ultra-small. As the angle approaches zero, the weak value $A_w$ becomes larger, which means the magnification strength becomes larger. But the inherent uncertainty is amplified correspondingly. The standard deviation of the experimentally measured signal is also contained in Fig. \ref{fig:03}(c). The undesired variations in practical measurement are mainly caused by a few factors: the power fluctuation of the light source, imperfect optical devices, air disturbance, etc.  This is also the case when the number of photons is too small, the influence of non-negligible background noise is more prominent in the data calculation. The contrast ratio's standard deviation ranges from $0.02$ to $0.05$. 

\begin{figure}[htbp]
\includegraphics[width=1\linewidth]{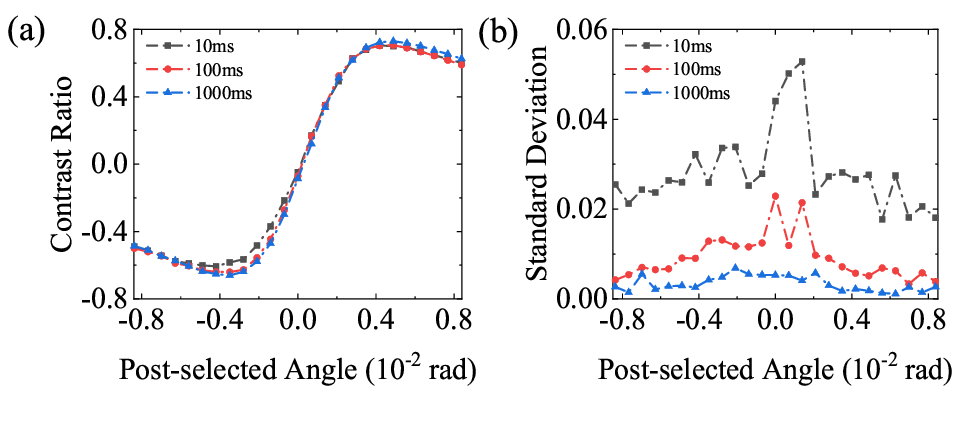}
\caption{The comparison of different integration times ($t$=10 ms, 100 ms, 1000 ms). (a) is the contrast ratio. The corresponding standard deviations are shown in (b).}
\label{fig:05}
\end{figure}

 In addition, with the increase in photon counts, the uncertainty of the contrast ratio is reduced as expected. As is in Fig. \ref{fig:04}, with the same integration time ($t$=10 ms), the higher the photon number $N$ is, the higher the precision is obtained. $N_{min}$=500 or 1500 refers to the minimum number of output photons ($N_++N_-$) received during a single measurement. The results in Fig. \ref{fig:04}(b) show that the precision is indeed increased due to the enhancement of signal power consistent with statistical analysis. Furthermore, three sets of measurements at different integration times ($t$=10 ms, 100 ms, 1000 ms) are compared in Fig. \ref{fig:05}. Illustrated by Fig. \ref{fig:05}(b), the detection signal's standard deviation of 1000ms is 10 times smaller than 10ms. It indicates that extending the integration time can also improve the precision. 
As clearly demonstrated by our experimental results, our new scheme can acquire a similar precision to CCD under the condition that incident power has a decrease of 4 orders of magnitude. 

\begin{figure}[htbp]
\includegraphics[width=1\linewidth]{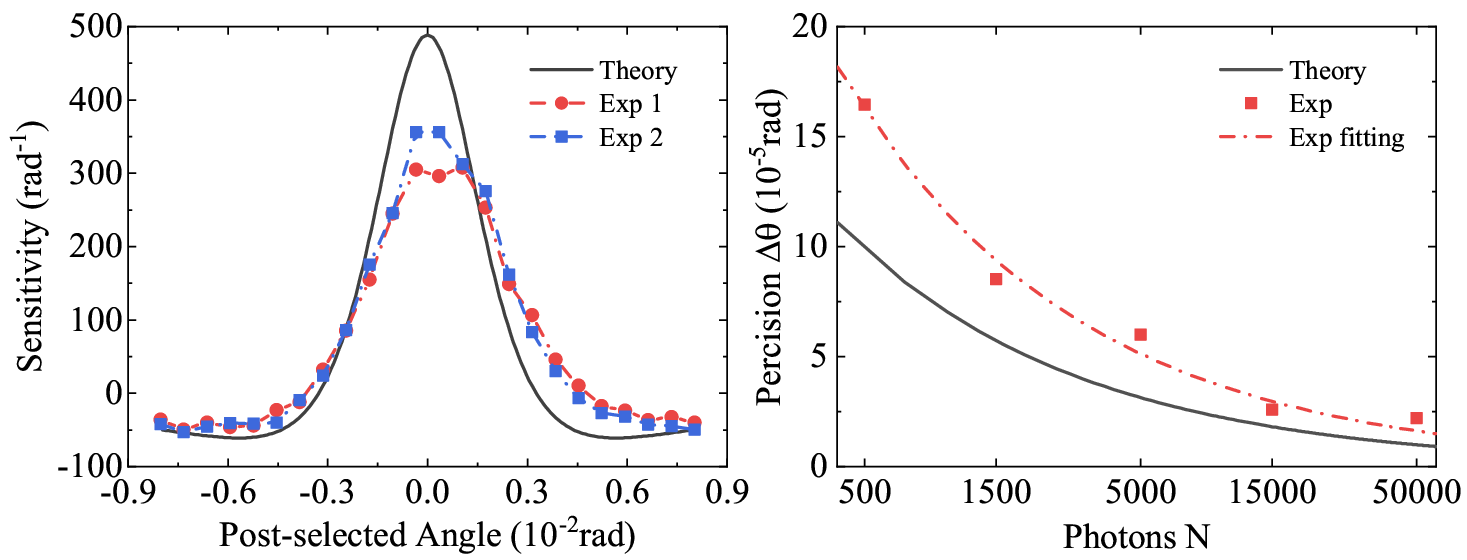}
\caption{(a) shows the sensitivity of the contrast ratio to the post-selected angle. The blue and red dots respectively denote the results with the number of minimum photons is 15000 or 50000 per measurement. Integration time is 100ms. (b) shows the mesured precision changing with the increasing receiving photons N. The solid black lines represent the theoretical value.}
\label{fig:06}
\end{figure}

Finally, the measured precision of the parameter $\theta$ depends on two key factors: the sensitivity of the measurement and the uncertainty of the signal. Therefore,the experimental results, as well as theoretical prediction curves, are shown in Fig. \ref{fig:06}. The theoretical formula for sensitivity ${\partial \langle\hat{\rm O}\rangle}/{\partial \theta}$ can be derived from Eq. (\ref{eq4}). It can be seen that the sensitivity of the two sets of data are in good agreement with the theoretical values in Fig. \ref{fig:06}(a). Meanwhile, concerning the precision changing with increasing photons N, when the number of photons ranges from 500 to 50000, the precisions improves as predicted by Eq. (\ref{eq8}). The highest precision obtained can reach $10^{-5}$ rad. As can be seen from the curves, the results are slightly inconsistency with the theoretical value, which is caused by the unsatisfactory disturbance. Nevertheless, these results indicate that even with limited resources our current measurement scheme can be employed to retrieve the original information efficiently. The data process is more simple and more convenient. Meanwhile, ultra-precise measurement can be achieved. Moreover, the lower integration time allows it to respond timely to dynamic signals, which can extend the applications.

{\it Discussion and conclusion.} 
A measurement strategy with a dimensionless pseudo-spin pointer is presented employing the generic framework of weak measurement.  With this new dimensionless pointer,  the moments of position can be obtained by counting the number of photons that arrived at instead of measuring the position directly. As a result, a highly sensitive and fast-responding detector SPAD can be employed at the detection port. Consequently, this makes the strategy naturally adapted to the measurement of weak signals in addition to its fast speed over the traditional measurement. With the current measurement strategy, the Cram\'er-Rao bound is shown to be saturated.  Our proof-of-principle experiment confirms the feasibility of our new method based on the dimensionless pointer to work even at extremely weak signals when the traditional method fails to work properly. The experimental results show that the precision of the strategy can achieve at least the same precision as the traditional method by explicitly measurement the position dependence of the signal with a sharp reduction of the cost time and hardware. It is expected that the idea of current measurement can be feasible for other quantum systems such as superconducting quantum interference devices and electron spin in solid device~\cite{wavefunction}.  

In this letter, we have confined our investigation to the case of independent measurements. As a result, the effect of quantum resources which has been shown can be exploited to go beyond the standard quantum limit is totally dismissed. However, according to Giovannetti {\it et al}.~\cite{q19}, the essential usage of entanglement will be at the stage of state preparation, and a classical measurement scheme will not ruin the quantum enhancement over the standard quantum limit. So it is interesting to study how to merge the current measurement strategy with existing quantum metrology protocols. Moreover, we only demonstrated the measurement of a two-state system with equal probability, which can be further extended to more general cases. Consequently, we believe that this detection technology can be widely used to detect various parameters and may have applications in broader fields. 

{\it Acknowledgments.}
This work is supported by the National Natural Science Foundation of China (Grant No. 11674234) and the Science Specialty Program of Sichuan University (Grant No. 2020SCUNL210). R.G. thanks the financial support of the "Fundamental Research Funds for the Central Universities".

\end{document}